\documentclass{aa}
\usepackage{epsfig,graphics}
\usepackage{wasysym}
\usepackage{txfonts}
\usepackage{natbib}
\bibpunct{(}{)}{;}{a}{}{,} 

\begin{document}

\title{Attempts to measure the magnetic field\\ of the pulsating B star \object{\boldmath{$\nu$}\,Eridani}}

\author{R.S. Schnerr\inst{1}
      \and E. Verdugo\inst{2}
      \and H.F. Henrichs\inst{1}
      \and C. Neiner\inst{3,4}
}

\institute{
      Astronomical Institute ``Anton Pannekoek'', University of Amsterdam, 
      Kruislaan 403, 1098 SJ Amsterdam, Netherlands
\and  European Space Astronomy Centre (ESAC), Research \& Scientific
      Support Department of ESA, Villafranca del Castillo, P.O. Box 50727, 28080 Madrid, Spain
\and  Institute for astronomy, KU Leuven, 3001 Leuven, Belgium
\and  GEPI, UMR\,8111 du CNRS, Observatoire de Paris-Meudon, 92195 Meudon Cedex, France
}

\offprints{R.S. Schnerr, 
\email{rschnerr@science.uva.nl}}

\date{Received 2 December 2005 / Accepted 7 March 2006}
\abstract{We report on attempts to measure the magnetic field of the pulsating B star
$\nu$ Eridani with the Musicos spectropolarimeter attached to the 2m telescope at the Pic du Midi,
France. This object is one of the most extensively studied stars for pulsation modes, and the existence of a magnetic field was suggested from the inequality of the frequency separations of a triplet in the stars' oscillation spectrum. We show that the inferred 5-10 kG field was not present during our observations, which cover about one year. We discuss the influence of the strong pulsations on the analysis of the magnetic field strength and set an upper limit to the effective longitudinal field strength and to the field strength for a dipolar configuration. We also find that the observed wind line variability is caused by the pulsations.

}
\titlerunning{Magnetic field measurements of \object{$\nu$\,Eri} }
\authorrunning{R.S. Schnerr et al.}
\maketitle

\section{Introduction} 

The B2III star \object{$\nu$ Eridani} (\object{HD 29248}, $V$ = 3.93) is known to show radial velocity variations for more than a century \citep{frost:1903}. It was found to be a multi-mode non-radial pulsator belonging to the class of $\beta$ Cephei variables with a main frequency of 5.76 c\,d$^{-1}$, identified as a $\ell=0$, p$_1$ mode. 
\citet{handler:2004a} and \citet{jerzykiewicz:2005} detected two independent low frequency, high-order g modes, demonstrating that the star also belongs to the class of Slowly Pulsating B (SPB) stars.
The star has the richest known oscillation spectrum of all $\beta$ Cephei stars. From a very extensive campaign \citep[see][hereafter Paper~I, II, III and IV]
{handler:2004a,aerts:2004,deridder:2004,jerzykiewicz:2005}, 34 photometric and 20 spectroscopic frequencies were detected, corresponding to 14 different pulsation frequencies. 

Among these 14, 12 are high-frequency modes, out of which 9 form three triplets, which are slightly asymmetric. The symmetric part is attributed to the effect of stellar rotation, whereas the asymmetric parts could be due to higher order rotational effects or due to a magnetic field. From line profile modeling \citet{smith:1983} derived $v\sin i$ $\approx$ 12 km\,s$^{-1}$ for the projected equatorial velocity. This value is consistent with a rotation period of 30 - 60 days as derived from the modeling of the splitting of the strongest triplet around 5.64  c\,d$^{-1}$, consisting of $\ell=1$ modes \citep[][hereafter DJ; Paper~I; Paper~II]{dziembowski:2003}.

DJ found that the asymmetry of this triplet, as measured from data of \citet{vanhoof:1961}, could only partly be explained by the quadratic effects of rotation \citep[see, e.g.,][]{saio:1981} and suggested that a strong magnetic dipole field of the order of 5-10 kG could explain this discrepancy. In a more recent analysis (Paper~I) the asymmetry was found to be a factor of 2 smaller than before, and \citet{pamyatnykh:2004} could entirely account for the asymmetry in terms of quadratic rotational effects. However, in Paper~III and IV the asymmetry was again found to be larger, and a second and third triplet were detected around 6.24  c\,d$^{-1}$ and 7.91  c\,d$^{-1}$. More advanced modeling of the different separations and asymmetries in all three triplets is still needed.

It is clear that an observational limit for the magnetic field strength will constrain such models, but until now no magnetic measurements of this star are available.
The specific prediction by DJ motivated us to observe this star with the Musicos spectropolarimeter at the T\'elescope Bernard Lyot (TBL, Pic du Midi, France) to search for the presence of a magnetic field, as the detection limit of this instrument is of the order of 100 G, which is far below the predicted value.
An additional argument to search for a magnetic field was the observed stellar wind variability as recorded 25 years earlier by the International Ultraviolet Explorer (IUE) satellite.

This paper describes our attempts to measure the magnetic field of \object{$\nu$~Eri} from spectropolarimetric observations.  We show the UV line variability as observed by the IUE satellite (Sect.\,\ref{subsec:iue}), describe how we interpret the signatures in our spectropolarimetric measurements as entirely due to the strong pulsations in this star rather than due to a magnetic field (Sect.\,\ref{subsec:model}) and set an upper limit to the field strength (Sect.\,\ref{sec:results}).

\section{Observations and data analysis} 
\subsection{IUE observations}
\label{subsec:iue}

\begin{figure}[t!bhp]
\caption[]{Ultraviolet \ion{C}{IV} line profile variability of $\nu$~Eri. The normalised flux in the upper panel is given in units of $10^{-9}$ erg~cm$^{-2}$s$^{-1}$\AA $^{-1}$. The lower panel display the ratio of the observed variance to the expected variance ($\sigma_{\mathrm{obs}}/\sigma_{\mathrm{exp}}$). The significant variations at red shifted wavelengths (around $\sim$100 km\,s$^{-1}$) are similar to those observed in other magnetic B stars, including the He strong and He weak stars. The spectra of $\nu$ Eri were corrected for the calculated radial velocity shift due to the pulsations.}
\label{iuec4}
\begin{center} 
\includegraphics[width=0.95\linewidth]{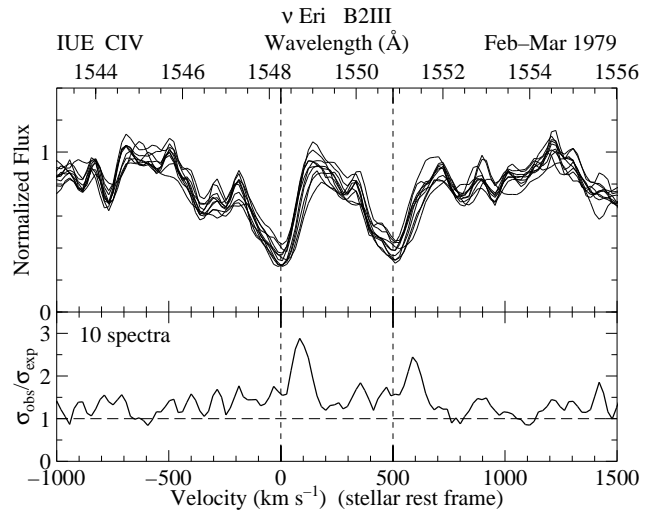}
\end{center}
\end{figure}

Specific behaviour of variable stellar wind lines belongs to the well-known indirect indicators of a magnetic field in early-type stars \citep[see][for a review]{henrichs:2003}. In Fig.~\ref{iuec4} we show the UV line profile changes in  $\nu$ Eri, as recorded with the IUE satellite in 1979.
The epochs of the observations and the radial velocity variations due to pulsation, which have been used to correct the spectra, are given in Table~\ref{iue}.

The temporal variance measures the ratio of the observed to the expected variability, and is very similar for the \ion{C}{IV} wind profiles in all magnetic B stars. We calculated the expected variability as a function of wavelength with a noise model as described by \citet{henrichs:1994}. The temporal variance spectrum of $\nu$ Eri (lower panel in Fig.~\ref{iuec4}) is very similar to that of a magnetic oblique rotator, such as $\zeta$~Cas \citep[a B2IV star, see][]{coralie:2003a}. Although only 10 high-resolution IUE spectra of $\nu$~Eri exist, nine of which were taken within one day and the other one month later, the peaks around +100 km~s$^{-1}$ in both doublet members are significant at the 3\,$\sigma$ level.

Although this variability is suggestive of the presence of a magnetic field in $\nu$~Eri, most of the variation is observed over one day. This period is consistent with known pulsation modes, which have periods between 0.126--2.312 days, rather than with the rotation period, expected to be of the order of months (see also Sect. \ref{subsec:uvdiscuss}).

\begin{figure}[t!bh]
\caption[]{UV continuum flux near \ion{C}{IV} measured with IUE (dots -- typical errors are 0.002) and the normalised U-band flux (line) deduced from the photometric pulsation analysis in Paper~I and III; the typical error range is indicated by the dashed lines. The timescales of the variability observed in the UV and in the U-band are very similar.}
\label{fig:iuelc}
\begin{center}
\includegraphics[width=0.97\linewidth,angle=0]{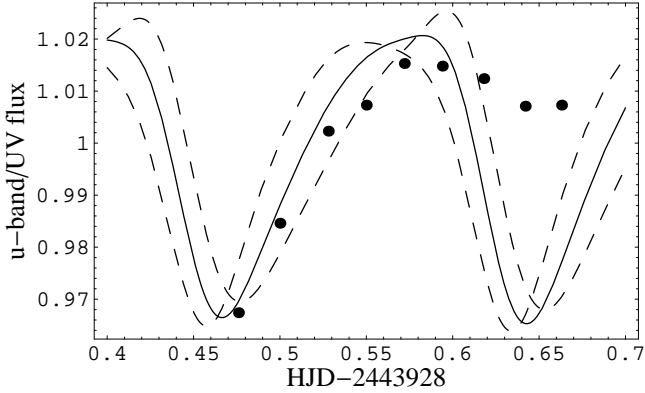}
\end{center}
\end{figure}

Before the temporal variance is constructed, the 10 spectra are normalised such that the equivalent widths summed over the wavelength bands [1465, 1510] \AA\ and [1575, 1605] \AA\ are equal to their average value.  These regions were selected to be free from stellar wind affected lines. The normalisation is necessary
because of the intensity variations due to the pulsations \citep[see, e.g.,][Paper~I]{porri:1994}. In Fig.~\ref{fig:iuelc} we show the inverse normalisation constants, which can be considered as a measure of the UV flux for the 9 spectra of Feb 23/24, 1979. The main timescale seen in the light curve is the same as found in several optical bands in Paper~I, and corresponds to that of the strongest pulsation mode detected in this star.

\begin{table}[b!thp]
\caption{Epochs of the IUE observations with applied radial velocity corrections and continuum ratios used to normalise the spectra (see text). The calculation of the radial velocities is based on Paper~III.}
\label{iue}
\begin{center}
\begin{tabular}{ll@{\ }ccrc}
\hline
\hline
Date & SWP  & HJD        & Exp. & \multicolumn{1}{c}{$v_{\rm rad}$} & \multicolumn{1}{c}{Normalisation}\\
1979 &      &($-$2443900)& s    & (km\,s$^{-1}$)& \multicolumn{1}{c}{near \ion{C}{IV} }\\
\hline
Feb. 23 & 4351 & 28.476 & 48.6&   29.2 & 1.0337\\
Feb. 24 & 4352 & 28.500 & 54.8&    8.8 & 1.0157\\
        & 4353 & 28.528 & 54.8&$-$19.2 & 0.9977\\
        & 4354 & 28.550 & 54.8&$-$25.5 & 0.9927\\
        & 4355 & 28.572 & 54.8&$-$18.3 & 0.9850\\
        & 4356 & 28.594 & 49.8& $-$2.3 & 0.9854\\
        & 4357 & 28.618 & 49.8&   17.6 & 0.9877\\
        & 4358 & 28.642 & 49.8&   26.8 & 0.9929\\
        & 4359 & 28.663 & 49.8&   21.5 & 0.9927\\
Mar. 29 & 4787 & 61.504 & 49.8& $-$5.7 & 1.0164\\
                          
\hline		    
\end{tabular}
\end{center}
\end{table}

\subsection{Spectropolarimetry}
\label{subsec:magnmeas}

{\small
\begin{table}[!b]
\caption{Journal of TBL observations, with epochs, exposure times and a comparison (O $-$ C) between the
measured radial velocities ($v_{\rm r, O}$) and predicted radial velocities ($v_{\rm r, C}$) based on the ephemeris and amplitudes of
Paper~III. Note that these values correspond reasonably well, provided that a constant difference of 14$\pm$1~km\,s$^{-1}$ for the system velocity is taken into account.}
\label{obs}
\begin{tabular}{l@{\ }l@{\ }ccrrrr}
\hline
\hline
Nr.& \multicolumn{1}{c}{Date}     & Mid HJD    & Exp. & \multicolumn{1}{c}{$v_{\rm r, O}$} & \multicolumn{1}{c}{	$v_{\rm r, C}$} & \multicolumn{1}{c}{O$-$C}\\
   &          & $-$2450000 & s & \multicolumn{1}{c}{km\,s$^{-1}$} & \multicolumn{1}{c}{km\,s$^{-1}$} & \multicolumn{1}{c}{km\,s$^{-1}$}\\
\hline
1a & 2003 Feb. 8 & 2679.300 & 600 & $-$9.7 & $-$20.5 & 10.8\\
1b &          & 2679.308 & 600 & $-$0.9 & $-$11.7 & 10.8\\
1c &          & 2679.316 & 600 &    7.4 &  $-$2.3 &  9.7\\
1d &          & 2679.323 & 600 &   16.7 &     6.1 & 10.6\\
2a & 2003 Oct. 25 & 2937.530 & 900 &   22.4 &     8.2 & 14.2\\
2b &          & 2937.548 & 900 &   31.2 &    16.6 & 14.6\\
2c &          & 2937.559 & 900 &   36.6 &    23.1 & 13.5\\
2d &          & 2937.570 & 900 &   43.4 &    31.5 & 11.9\\
3a & 2004 Feb. 4 & 3040.452 & 750 &   19.5 &     1.6 & 17.9\\
3b &          & 3040.461 & 750 &   14.3 &  $-$0.1 & 14.4\\
3c &          & 3040.471 & 750 &   12.3 &     1.6 & 10.7\\
3d &          & 3040.480 & 750 &   15.4 &     2.5 & 12.9\\
4a & 2004 Feb. 6 & 3042.412 & 750 & $-$3.3 & $-$16.6 & 13.3\\
4b &          & 3042.421 & 750 & $-$6.6 & $-$19.0 & 12.4\\
4c &          & 3042.430 & 750 & $-$7.6 & $-$20.7 & 13.1\\
4d &          & 3042.439 & 750 & $-$6.8 & $-$19.5 & 12.7\\
5a &  2004 Feb. 8 & 3044.445 & 750 &   50.5 &    27.8 & 22.7\\
5b &          & 3044.454 & 750 &   45.5 &    21.3 & 24.2\\
5c &          & 3044.464 & 750 &   32.3 &    10.1 & 22.2\\
5d &          & 3044.473 & 750 &   12.4 &  $-$3.5 & 15.9\\
6a & 2004 Feb. 10 & 3046.289 & 750 &    0.7 & $-$10.5 & 11.2\\
6b &          & 3046.298 & 750 &    8.3 &  $-$1.4 &  9.7\\
6c &          & 3046.308 & 750 &   15.9 &     8.0 &  7.9\\
6d &          & 3046.317 & 750 &   23.4 &    14.5 &  8.9\\
7a & 2004 Feb. 12 & 3048.413 & 750 &   27.6 &    14.2 & 13.4\\
7b &          & 3048.422 & 750 &   29.5 &    14.4 & 15.1\\
7c &          & 3048.431 & 750 &   30.9 &    14.9 & 16.0\\
7d &          & 3048.441 & 750 &   33.5 &    18.2 & 15.3\\
8a & 2004 Feb. 14 & 3050.272 & 750 &    9.8 &  $-$6.6 & 16.4\\
8b &          & 3050.282 & 750 &   13.6 &  $-$1.9 & 15.5\\
8c &          & 3050.291 & 750 &   18.3 &     3.6 & 14.7\\
8d &          & 3050.300 & 750 &   23.2 &     8.8 & 14.4\\
\hline
\end{tabular}
\end{table}
}

The magnetic field measurements were carried out with the Musicos spectropolarimeter attached to the 2m TBL at the Pic du Midi, France. We obtained 32 spectra of $\nu$ Eri between 8 Feb. 2003 and 14 Feb. 2004 (see Table~\ref{obs}) from which circularly polarised (Stokes $V$) and unpolarised spectra (Stokes $I$) are calculated. 
The technique to carry out
high-precision magnetic measurements with this instrument is extensively described by
\citet{donati:1997} and \citet{wade:2000}. Each set of four subexposures was taken in the usual $\lambda/4$-plate position sequence q1, q3, q3, q1, corresponding to $\pm 45^{\circ}$ angles. 
We used the dedicated ESpRIT data reduction package \citep{donati:1997} for the optimal extraction of the \'{e}chelle spectra and to obtain Stokes $I$ and $V$ spectra. 
We also calculate a Null polarisation, called Stokes $N$, which represent the pollution by non-magnetic effects and should be null for a perfect measurement.
The package includes a Least-Squares Deconvolution (LSD) routine to calculate a normalised average Stokes $I$ line profile and corresponding Stokes $V$ and $N$ line profiles of all available spectral lines (we used 107-108 lines). If a magnetic field is present it will result in a typical Zeeman signature in the average Stokes $V$ profile, from which the effective longitudinal component of the stellar magnetic field can be determined (see Sect.~\ref{magfieldmeas}).

\subsubsection{Polarisation signatures}

Stokes $V$ and $N$ spectra were calculated using the standard equations:
\begin{equation}
\frac{V}{I}=\frac{R_V-1}{R_V+1}; \hspace{0.5cm} \frac{N}{I}=\frac{R_N-1}{R_N+1}
\end{equation}
where
\begin{equation}
R^4_V = \frac{I_{1\perp} \cdot I_{3\parallel}}{I_{1\parallel} \cdot I_{3\perp}} \cdot \frac{I_{2\parallel} \cdot I_{4\perp}}{I_{2\perp} \cdot I_{4\parallel}} ~~\mathrm{and}~~
R^4_N = \frac{I_{1\perp} \cdot I_{3\parallel}}{I_{1\parallel} \cdot I_{3\perp}} \cdot \frac{I_{2\perp} \cdot I_{4\parallel}}{I_{2\parallel} \cdot I_{4\perp}}.
\label{eq:stokesVN}
\end{equation}
The symbols $I_{k\perp}$ and $I_{k\parallel}$ represent the perpendicular and parallel beams
emerging from the beam splitter of subexposure $k$, respectively.  The $\lambda$/4-plate orientations during
the two subexposure pairs \{$1, 4$\} and \{$2, 3$\} are perpendicular to each other.

\begin{figure}[!tbhp]
\caption[]{LSD results for $\nu$ Eri on 8 Feb. 2003. The average intensity line profile (bottom), the Stokes $V$ profile (middle) and the Stokes $N$ profile (top) are shown. $V$ and $N$ are shifted up by 1.05 and 1.10 respectively for display purposes. Note the signature in $V$ which is typical for a magnetic field, but the presence of a signature in $N$, generated by the pulsations, indicates that the Stokes $V$ profile could be affected (see text).}
\label{stokesv}
\begin{center}
\includegraphics[height=0.9\linewidth,angle=-90]{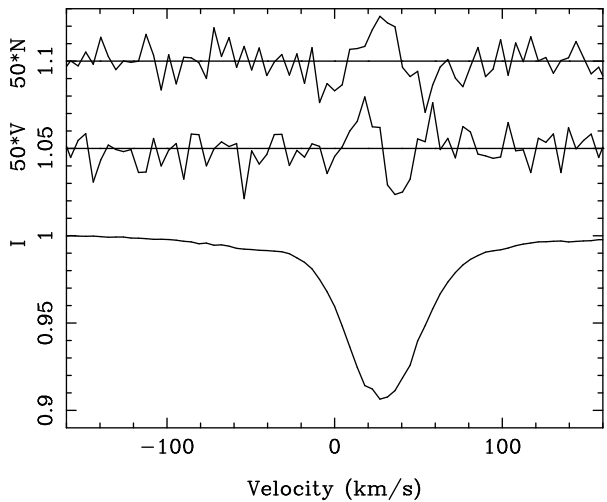}
\end{center}
\end{figure}

Although no significant magnetic fields are detected (see Table~\ref{table:results}), in several cases the average Stokes $V$ profiles (see Fig.~\ref{stokesv} and \ref{example_profiles}) show a significant signature which could be interpreted as due to a magnetic field. However, significant signatures are also visible in the corresponding Stokes $N$ profile, which makes a magnetic interpretation questionable, especially because of the large changes in the line profile due to pulsations during the 4 subexposures.

We now consider the origin of the signatures. In an ideal instrumental setup, pulsations would not create signatures in Stokes $V$ (or $N$) since for each spectrum both right-handed circular and left-handed circular polarisation spectra are recorded simultaneously, and different line shapes between the four subexposures of one magnetic field measurement should cancel exactly (see Eq.~\ref{eq:stokesVN}). The practical reason why they do not cancel is that the two spectra of one subexposure partially follow a different light path and are recorded on different pixels of the CCD.  This is why usually at least 2 (and often 4) subexposures are used. As a consequence the two spectra may have a different intensity level and slight differences in wavelength calibration, on average typically several hundred m~s$^{-1}$ \citep{semel:1993,donati:1997}. Such inevitable inaccuracies cause the different line shapes between the different subexposures to appear in the resulting Stokes $V$ spectrum. For stars that do not show strong changes in line shape this problem does not occur, because the differences between the two spectra of one subexposure are corrected by the next subexposure where the opposite circular polarization state is recorded through the same light path, on the same pixels and with the same wavelength calibration. In the following section we closely examine the effect of these differences.

\subsection{Modeling the Stokes $V$ and $N$ profiles}

\begin{figure*}[!tbhp]
\caption[]{Stokes $I$, $N$ and $V$ profiles of observations 1 ({\sl left}), 2 ({\sl middle}) and 6 ({\sl right}). These profiles were selected to illustrate the effect of the pulsation. In the Stokes $I$ plots ({\sl top}), the solid lines represent the model fits. The dashed lines are the four exposures that were used for one magnetic field measurement. In the plots of the Stokes $N$ ({\sl middle}) and $V$ ({\sl bottom}, in \permil), the dashed line represents the observations, the thin solid line the model, and the thick solid line the corrected observations after subtracting the model to correct for the effect of the pulsations. Note that the corrected Stokes $N$ profiles are all consistent with zero, which indicates that the assumption of a constant velocity shift between the two beams is sufficient to explain the signatures in Stokes $N$ for pulsation-affected profiles. The limits of [$-$0.04\%,+0.04\%] adopted in Sect.\,\ref{section:bul} as the maximum amplitude of any undetected, broad, magnetic polarisation signatures, are indicated by the dashed lines in the bottom left plot.}
\label{example_profiles}
\begin{minipage}[t]{0.04\linewidth}
\rotatebox[]{90}{Flux (\permil)  \hspace{4cm} Norm. Flux \hspace{1cm}}
\end{minipage}
\begin{minipage}[t]{0.31\linewidth}
\begin{flushleft}
\includegraphics[width=0.95\linewidth]{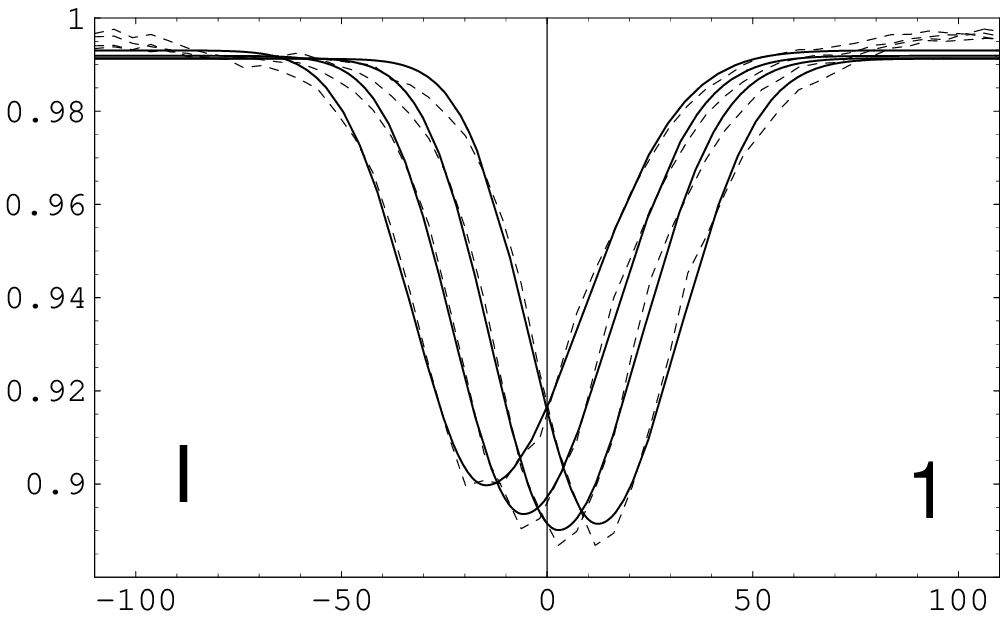}
\includegraphics[width=0.95\linewidth]{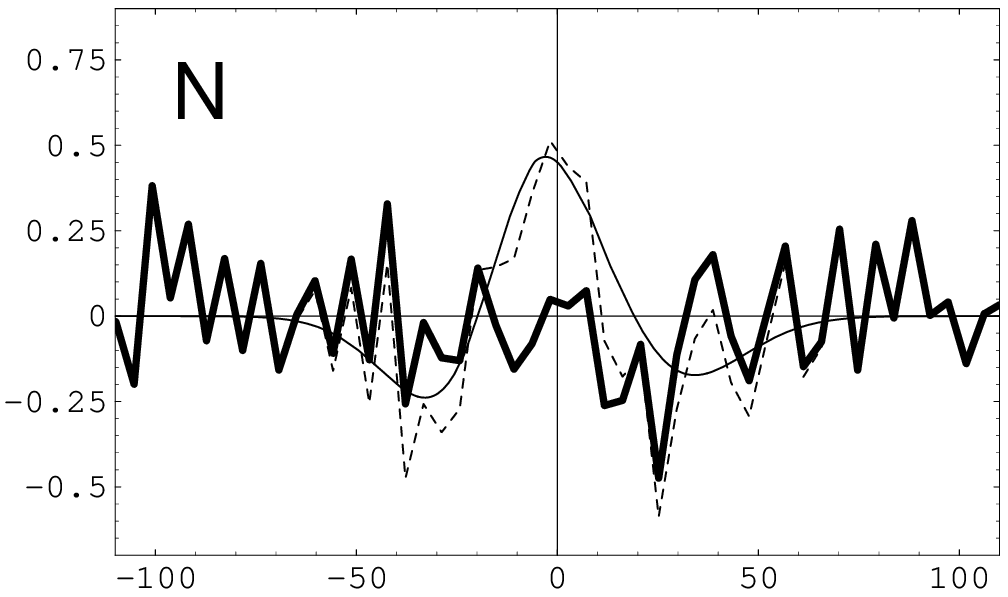}
\includegraphics[width=0.95\linewidth]{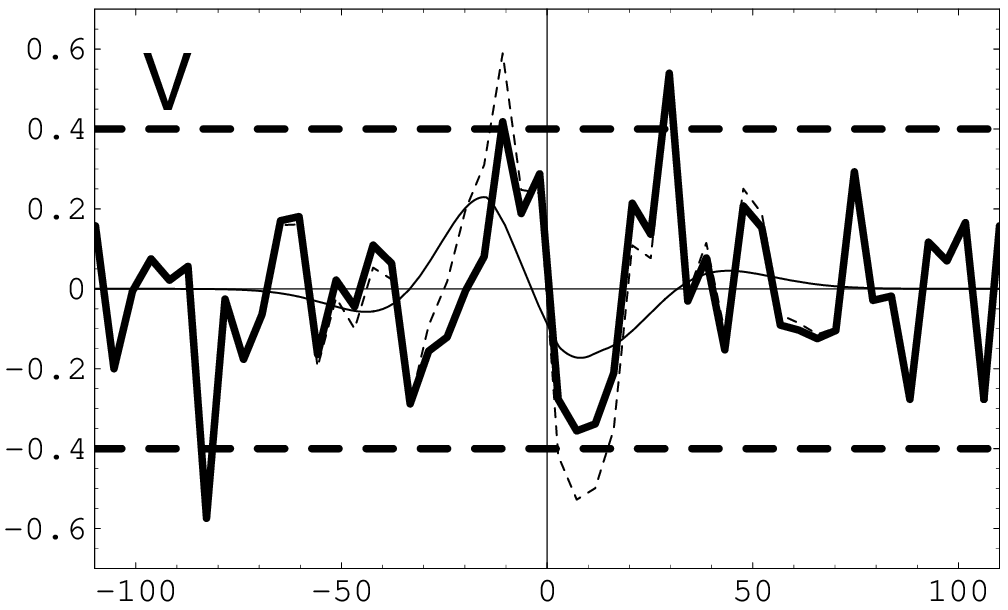}
\end{flushleft}
\end{minipage}
\begin{minipage}[t]{0.31\linewidth}
\begin{center}
\includegraphics[width=0.95\linewidth]{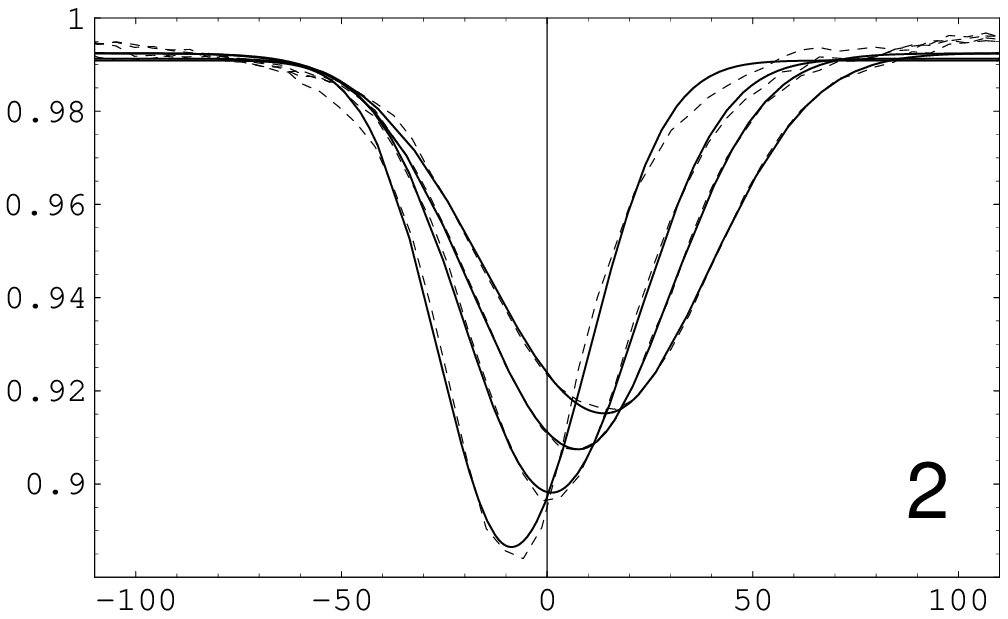}
\includegraphics[width=0.95\linewidth]{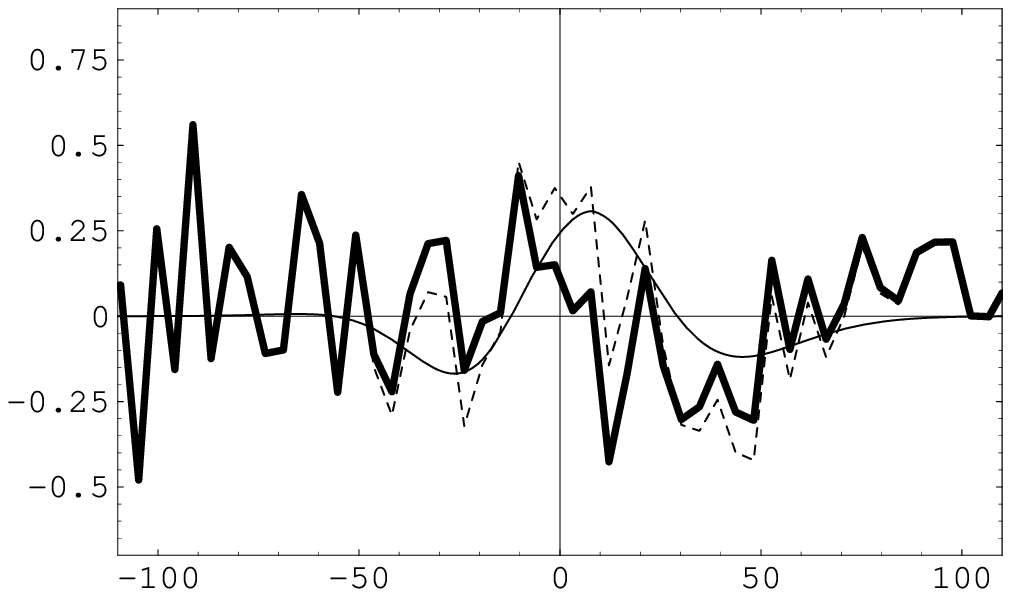}
\includegraphics[width=0.95\linewidth]{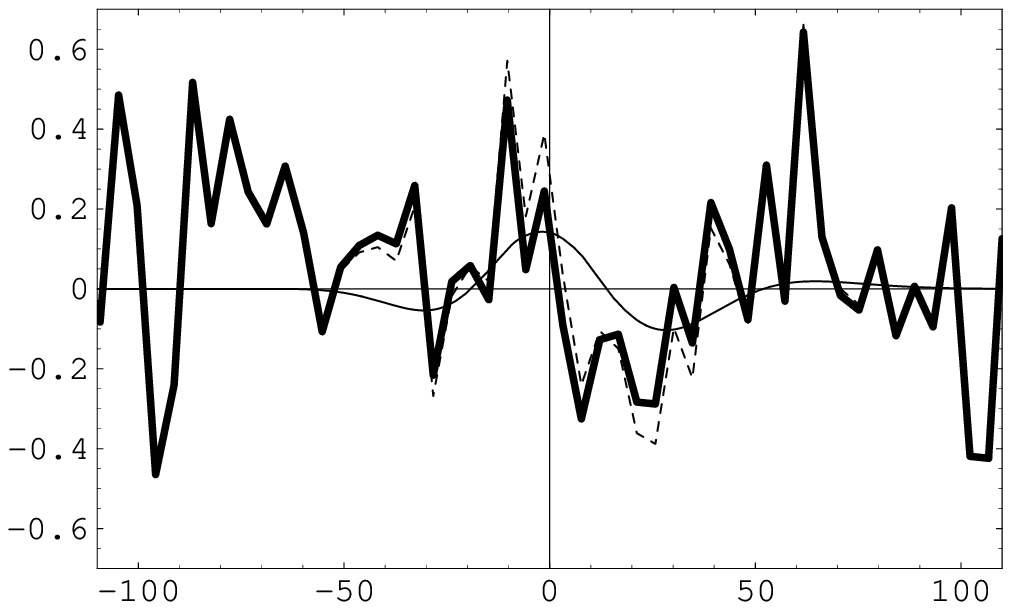}
Velocity (km/s)
\end{center}
\end{minipage}
\begin{minipage}[t]{0.31\linewidth}
\begin{flushright}
\includegraphics[width=0.95\linewidth]{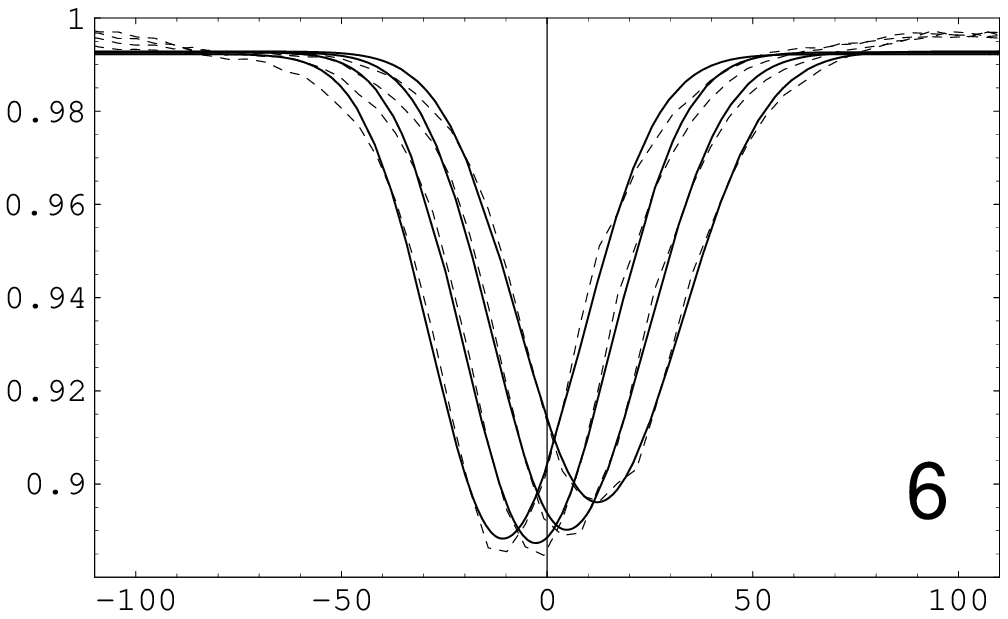}
\includegraphics[width=0.95\linewidth]{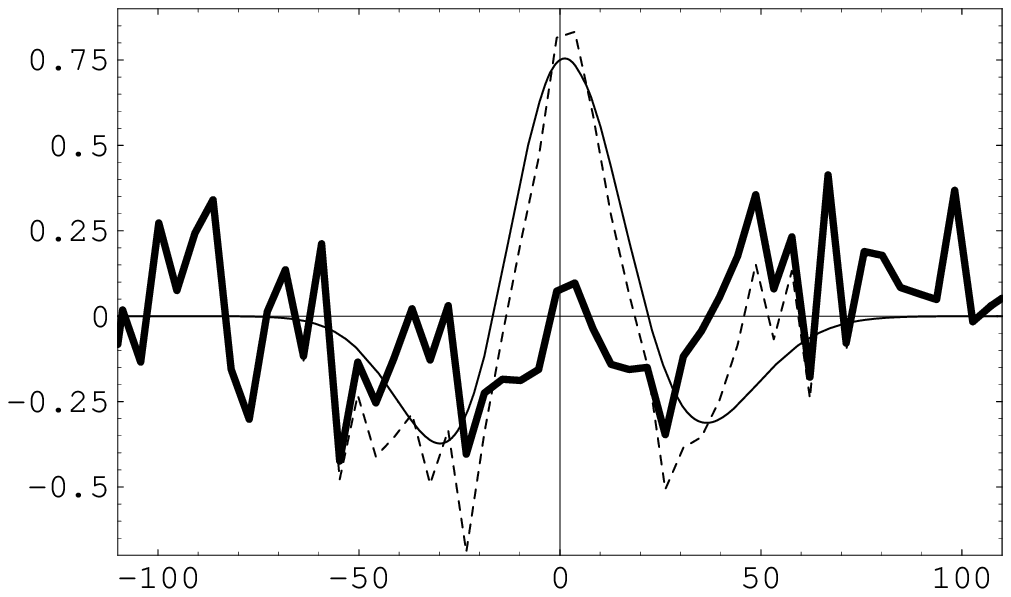}
\includegraphics[width=0.95\linewidth]{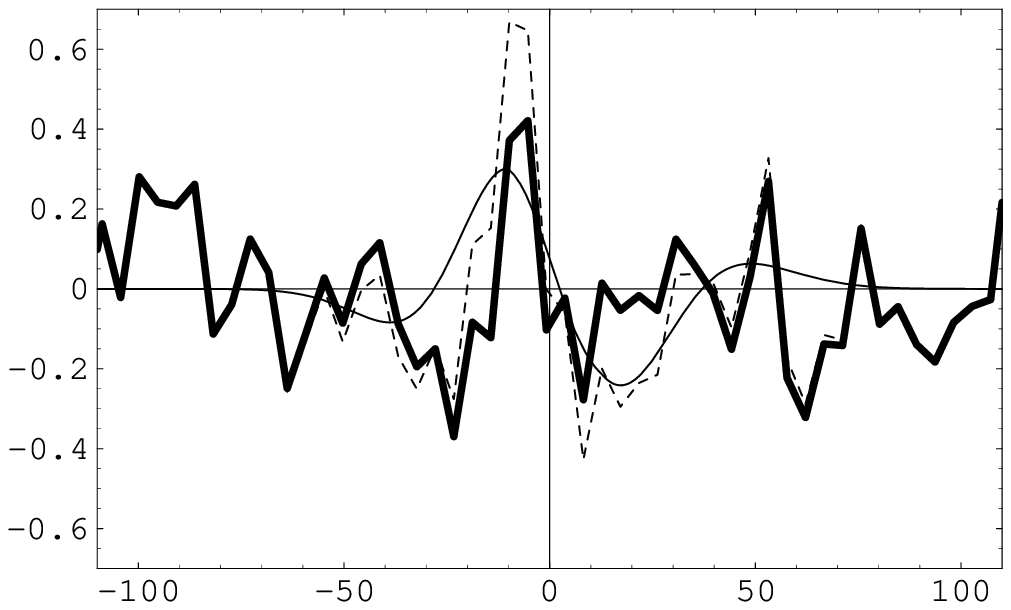}
\end{flushright}
\end{minipage}
\end{figure*} 

\label{subsec:model}
To investigate the effect of pulsations on the signatures in the Stokes $V$ profile we developed a model that predicts Stokes $V$ and $N$ signatures from the Stokes $I$ profiles of all four subexposures. For each subexposure we first fit the average line profile, resulting from the LSD method, with the following function:

\begin{equation} I(v,v_{\mathrm{rad}},c_1, c_2, c_3, c_4)=c_1\,\mathrm{exp}[-f(v,v_{\mathrm{rad}},c_2,c_3,c_4)], \label{eq:lineprof} \end{equation}
in which
\begin{displaymath} f(v,v_{\mathrm{rad}},c_2,c_3,c_4)=c_2\,\mathrm{exp}\left[-\left(\frac{v-v_{\mathrm{rad}}}{[1+c_3 \mathrm{Sign}(v-v_{\mathrm{rad}})]c_4}\right)^2\right]. \end{displaymath}

In this equation the only variable is the velocity parameter $v$, whereas the five constants are $v_{\mathrm{rad}}$ the radial velocity of the line, $c_1$ the continuum level, $c_2$ the minimum intensity level relative to $c_1$, $c_3$ the asymmetry parameter and $c_4$ the full line width. These five constants are determined for each subexposure by a least-squares best fit procedure.

To model the error in the wavelength calibration between the two beams of one subexposure, we characterise the typical wavelength shift between the two beams with a velocity parameter $v_{\rm{shift}}$. From the resulting 8 line profiles (4 subexposures $\times$ 2 beams) we calculate Stokes $V$ and $N$ profiles, with a fixed value for $v_{\rm{shift}}$ to be determined by the fit. It is important to note that such an offset in wavelength calibration has a different effect than the presence of a magnetic field. A magnetic field would cause the spectrum in one circular polarisation state to be shifted relative to the other due to the Zeeman splitting; this shift is of opposite sign in the q1 and q3 exposures due to the switching of the two polarisation states between the subsequent subexposures. In our case the shift is the same in both q1 and q3 subexposures because this parameter is related to the beams themselves.

We determine the parameter $v_{\rm{shift}}$ by a minimum $\chi^2$ fit of the calculated Stokes $N$ profiles from our model to the measured Stokes $N$ profiles. The results can be found in Table~\ref{table:results}. To check whether these values are realistic, we have extracted several ThAr exposures in the same way as we do for our science exposures. Since ThAr exposures are used to wavelength calibrate the spectra, we would expect similar inaccuracies between the two beams in these exposures as for our science exposures. We indeed find shifts in velocity between spectral lines in the two beams, varying from $\sim-$3 to +1 km~s$^{-1}$, explaining the average shifts found.

In Fig.~\ref{example_profiles} we show the Stokes $I$, $N$ and $V$ profiles of the three observations which were subjected to the strongest pulsations. It is clear that the features in the Stokes $N$ profile can be fairly well reproduced with our simple model, which contains only one free parameter. The signatures in the Stokes $V$ that are predicted by our model have a somewhat smaller amplitude and appear to be slightly broader than the observed profiles, but the overall shape is very similar. The broadness of the modeled profiles is probably due to the fact that we use only one parameter, $v_{\rm{shift}}$, to characterise the shift of the average line profile, while in reality there is a distribution for all the different lines. Furthermore, at some phases the line profiles show extended wings which are not represented in our model, and resulted in slightly broader model fits.

With this modeled contribution to the Stokes $V$ profile we can quantitatively determine the spurious effect on the magnetic field determination as will be done in Sect.\,\ref{magfieldmeas}.

\section{Results and discussion}
\label{sec:results}

\subsection{UV variability}
\label{subsec:uvdiscuss}
The variability observed in the wind-lines of $\nu$ Eri looks similar to that observed for the magnetic early B-type stars \object{$\beta$~Cep}, \object{$\zeta$~Cas}, \object{V2052~Oph} and \object{$\omega$~Ori} \citep{henrichs:2000a,coralie:2003a,coralie:2003c,coralie:2003b}. However, since the timescale of this variability is comparable to the timescale of the pulsations rather than the rotation period (1--2 months), one can conclude that it is the pulsations that are responsible for the observed variability. The observed range of wind variability estimated from the variance is of order 50--100 km s$^{-1}$ (Fig.~\ref{iuec4}), which is in the same range as the expected range from the pulsations in this star. This is similar to what is observed in the magnetic stars, implying that the same low-velocity part of the stellar wind is affected, even though different mechanisms are involved. This phenomenon, where the low-velocity part of the stellar wind is influenced by strong pulsations has previously been observed in strong pulsators, such as \object{BW~Vul} \citep{burger:1982,smith:2003}.

\begin{table*}[!tbhp]
\caption{Results of the TBL magnetic field measurements. The signal to noise ratio per pixel (S/N) was measured around 550 nm in the Stokes $V$ spectrum (order 108). The HJD was calculated halfway the four subexposures used for each magnetic measurement. Measurements shown are the effective magnetic field as measured from the Stokes V profiles before and after correcting for the pulsations ($B_{\mathrm{eff}}$ and $B_{\mathrm{corr}}$ respectively) and similarly for the values for $N$. The last column gives the best-fit velocity shift between the two beams.}
\label{table:results}
\begin{center}
\begin{tabular}{clrlcrrrrrrr@{$\pm$}l}
\hline
\hline
Obs.& \multicolumn{1}{c}{Date}  & \multicolumn{1}{c}{HJD (mid obs.)} & S/N & Range       & $B_{\mathrm{eff}}$ &  $B_{\mathrm{corr}}$ & $\sigma_B$ & $N_{\mathrm{eff}}$ & $N_{\mathrm{corr}}$ & $\sigma_\mathrm{N}$ & \multicolumn{2}{c}{Velocity shift}\\
    &                           & \multicolumn{1}{c}{$-$2452000}     &     & (km~s$^{-1}$)      & (G)       & (G)         & (G)	    & (G)       & (G)        & (G)        & \multicolumn{2}{c}{(km~s$^{-1}$)}\\
\hline
1 & 2003 Feb. 8   		&  679.3090  	  		     & 560 &  [$-$63,68] & $-$5      &  $-$7	   & 42 	& 20	    & 25	 & 41	      & $-$0.42&0.08 \\
2 & 2003 Oct. 25  		&  937.5507  	  		     & 430 &  [$-$66,78] &    5      &  2	   & 61 	& 63	    & 61	 & 62	      & $-$0.43&0.16 \\
3 & 2004 Feb. 4   		& 1040.4632  	  		     & 120 &  [$-$86,92] &  437      &  419	   & 314	& 57	    & 78	 & 316        & $-$2.71&3.75 \\
4 & 2004 Feb. 6   		& 1042.4229  	  		     & 590 &  [$-$65,69] &   59      &  59	   & 39 	& 26	    & 26	 & 36	      &    0.15&0.36 \\
5 & 2004 Feb. 8   		& 1044.4563  	  		     & 430 & [$-$137,105]&  163      &  153	   & 127	& 77	    & 71	 & 125        & $-$0.43&0.36 \\
6 & 2004 Feb. 10  		& 1046.3001  	  		     & 560 &  [$-$59,65] &   11      &  6	   & 39 	& $-$75     & $-$74	 & 38	      & $-$0.71&0.08 \\
7 & 2004 Feb. 12  		& 1048.4243  	  		     & 510 &  [$-$65,71] &   36      &  36	   & 46 	& $-$2      & $-$1	 & 44	      & $-$0.43&0.36 \\
8 & 2004 Feb. 14  		& 1050.2834  	  		     & 670 &  [$-$61,65] & $-$9      &  $-$11	   & 31 	& $-$6      & $-$6	 & 29	      & $-$0.57&0.12 \\
\hline
\end{tabular}
\end{center}
\end{table*}

\subsection{Magnetic field measurements}
\label{magfieldmeas}
The longitudinal component of the magnetic field averaged over the stellar disc in Gauss, is, as usual, calculated as (in velocity space):
\begin{equation}
\label{V_moment}
B_{\mathrm{eff}}=2.14 \times 10^{11} \frac{\int v\,V(v)\,\mathrm{d}v}{\lambda\,g\,c \int [1-I(v)] \mathrm{d}v},
\label{eq:bfield}
\end{equation}
where $\lambda$ is the average wavelength of the used lines in nm, $g$ is their averaged Land\'{e} factor, and $c$ the speed of light in cm/s. To estimate the strength of the signal in $N$, we calculated $N_{\rm{eff}}$ from the Stokes $N$ profile analogous to Eq.~\ref{V_moment}.

In Table~\ref{table:results} we show the effective magnetic field strength as measured from the observations both before and after subtracting the modeled signature of the pulsations. The integration ranges used are set at two times the width of the fitted line profile as determined from Eq.~\ref{eq:lineprof} (see also Table~\ref{table:results}). In general, the lower and upper limits of the integration are different due to the varying shape of the line profile. The difference in $B_{\rm{eff}}$ and $N_{\rm{eff}}$ with and without the correction for pulsations is quite small. This is because the signatures created by the pulsations are almost symmetric, and hence do not much influence the first moments that determine these quantities.

No magnetic field has been detected.
However, for some observations significant signatures in Stokes $V$ are found. This could indicate that although the effective longitudinal magnetic field strength is zero, there is still evidence for the presence of a magnetic field. Using our simple model, we have shown that we are able to model the signatures found in Stokes $N$ very well, and at the same time predict Stokes $V$ profiles that are very similar in shape to the measured ones. From this we conclude that the profiles we detect in Stokes $N$ and $V$ are the result of the combined effects of the pulsations and inaccuracies in wavelength calibration that were not removed by our imperfect modeling of these effects.

\subsection{Constraining the magnetic field}
\label{section:bul}

\begin{figure}[!tbhp]
\caption[]{For a given magnetic field strength, the amplitude of the Stokes $V$ signature depends mainly on $\alpha$, the angle between the rotation axis projected onto the plane of the sky and the magnetic axis.}
\label{fig:alpha}
\begin{center} 
\includegraphics[width=0.9\linewidth, clip]{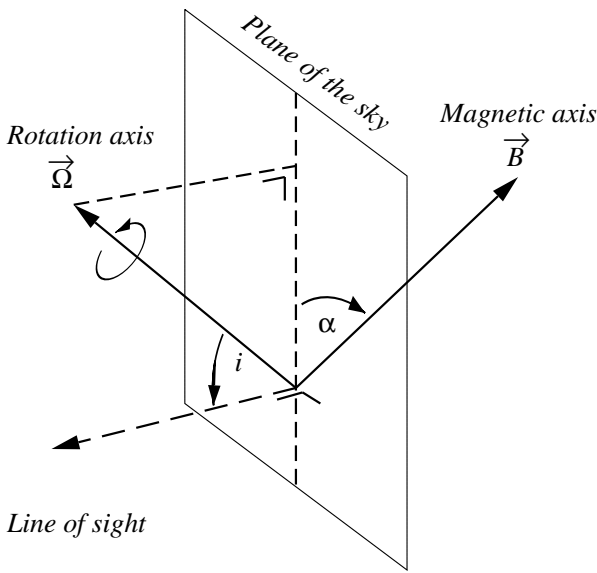}
\end{center}
\end{figure}

\begin{figure*}[!tbhp]
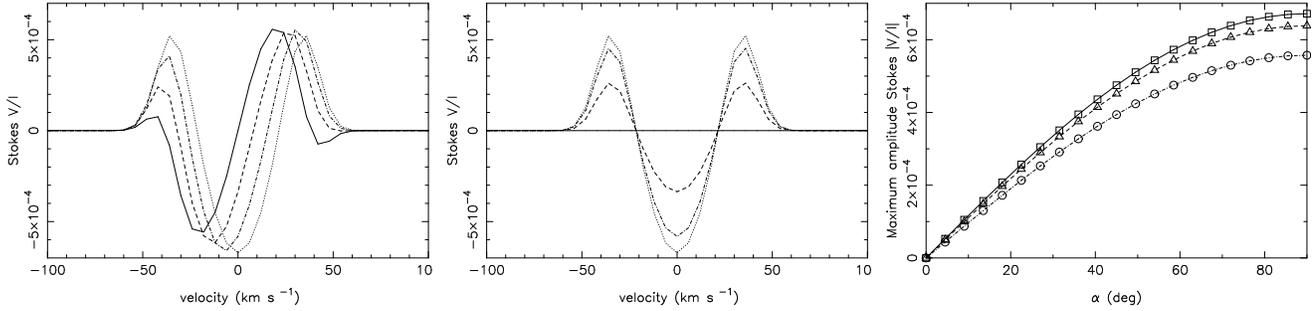

\caption[]{Results of model calculations of Stokes $V$ profiles for a star with a dipolar magnetic field with a polar strength of 300 G. The {\sl left} plot shows profiles for an inclination of $i$=0$^\circ$, an angle between the rotation and magnetic axis of $\beta=90^\circ$ and for rotation angle $\phi=0^\circ$ (magnetic axis pointing towards observer, maximum $B_{\mathrm{eff}}$ -- solid line), $\phi=30^\circ$ (dashed line), $\phi=60^\circ$ (dashed-dotted line) and $\phi=90^\circ$ (zero $B_{\mathrm{eff}}$ -- dotted line). Although the shape of the signature and $B_{\mathrm{eff}}$ vary, the maximum amplitude remains approximately constant.
The {\sl middle} plot shows the profiles for $i$=0$^\circ$, $\phi=90^\circ$ and $\beta=0^\circ$ (solid line), $\beta=30^\circ$ (dashed line), $\beta=60^\circ$ (dashed-dotted line) and $\beta=90^\circ$ (dotted line). On the {\sl right} we show the maximum amplitude of the Stokes $V$ profile vs. $\alpha$ for $\phi=90^\circ$  (squares), $\phi=45^\circ$ (triangles) and $\phi=0^\circ$ (circles). The lines represent $\sin(\alpha)$ normalised to $\alpha=90^\circ$ (with $i$=0$^\circ$, $\beta=\alpha$). The maximum amplitude roughly scales with $\sin(\alpha)$ with a slight dependency on the orientation (as can also be seen in the left plot). }
\label{fig:signatures}
\begin{minipage}{0.32\linewidth}
\begin{flushleft}
\includegraphics[height=\linewidth,angle=-90, trim=80 10 10 80, clip]{4635fig6a.ps}%
\end{flushleft}%
\end{minipage}
\begin{minipage}{0.32\linewidth}
\begin{center}%
\includegraphics[height=\linewidth,angle=-90, trim=80 10 10 80, clip]{4635fig6b.ps}
\end{center}
\end{minipage}
\begin{minipage}{0.32\linewidth}
\begin{flushright}
\includegraphics[height=\linewidth,angle=-90, trim=80 10 10 80, clip]{4635fig6c.ps}
\end{flushright}
\end{minipage}
\end{figure*}

Constraining the magnetic field of a star from Stokes $V$ profiles is not straightforward. A Stokes $V$ profile is only sensitive to the line-of-sight component of the field and a light-intensity weighted average over the visible stellar disc is involved, similar to the formation of a line profile. Since one also has to account for the rotational Doppler shifts of the lines, it will be clear that Stokes $V$ profiles can be very different for stars with the same (polar) field strength $B_{\rm p}$ and $v \sin i$, even for a simple dipolar configuration.

To determine the constraints on the polar field strength from our spectropolarimetry we used a model that calculates Stokes $V$ profiles from the Zeeman splitting of an absorption line. In this simple model a spectral line at rest wavelength $\lambda_0$ is split into two Zeeman components with wavelength $\lambda_0-\Delta\lambda_H$ and $\lambda_0+\Delta\lambda_H$, where $\Delta\lambda_H$ is the typical wavelength shift corresponding to the local line-of-sight component of the magnetic field \citep{mathys:1989}. To determine the final profile we integrate over the visible stellar disc, using a limb darkening constant $\epsilon=0.3$ \citep{gray:1992}, $v \sin i=40$ km~s$^{-1}$, and an intrinsic line width of 10 km~s$^{-1}$. This high value for $v \sin i$ is required to reproduce the average line profile, which is broadened by pulsations and the averaging process. We checked this model by reproducing Stokes $V$ profiles for $\beta$ Cep which has a polar field strength of 360 G \citep{henrichs:2000a,donati:2001}.

From this model we find that for a given field strength and $v \sin i$ the amplitude of the Stokes $V$ profile mainly depends on the angle between the rotation axis projected onto the plane of the sky and the magnetic axis (the angle $\alpha$ in Fig.~\ref{fig:alpha}). Although the {\it shape} of the profile and $B_{\mathrm{eff}}$ depend on whether the magnetic axis is pointing towards or away from us (maximum and minimum $B_{\mathrm{eff}}$) or lies in the plane of the sky ($B_{\mathrm{eff}}=0$), the amplitude of the profile is rather independent of this. Example profiles and the dependence of the maximum amplitude on $\alpha$ are shown in Fig.~\ref{fig:signatures}. The maximum amplitude approximately scales with $|\sin \alpha|$.

\begin{figure}[!bthp]
\caption[]{Upper limit to the magnetic field strength at the magnetic pole as a function of the angle $\alpha$.}
\label{fig:upperlimits}
\begin{center} 
\includegraphics[height=0.9\linewidth,angle=-90]{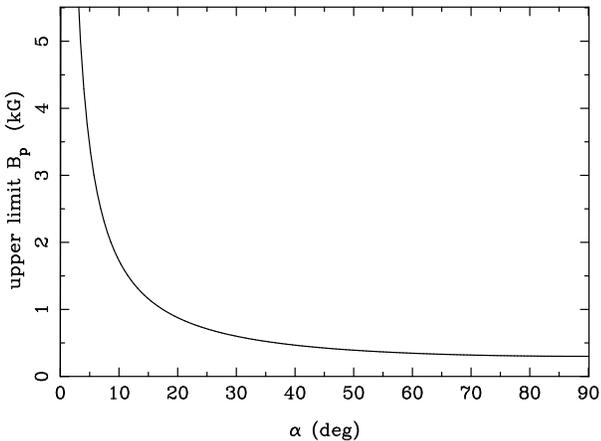}
\end{center}
\end{figure}

For our observations (except for nr. 3, which has a very low S/N), magnetic polarisation signatures in Stokes $V$ with an amplitude larger than approximately 0.04\% would have been detected (see Fig.~\ref{example_profiles}). With our model it is possible to constrain the strength of a dipolar magnetic field for a given $\alpha$. For a maximum amplitude of 0.04\%, we find that the upper limit on the polar field strength is: $B_{\rm p} \lesssim 300 {\rm \,[G]}/\sin\alpha$ (see Fig.~\ref{fig:upperlimits}). To hide the predicted field of $B_{\rm p}\geq 5$ kG, $\alpha$ would have to be smaller than about 3.5$^\circ$. So the angle between the magnetic field axis and the projected rotation axis would have to be smaller than this 3.5$^\circ$, for all observations.

Generally (unless the rotation axis lies exactly in the plane of the sky) $\alpha$ depends on the rotational phase. Our observations of February 2004 cover a period of 10 days, which, with an estimated period of $\nu$ Eri of 1--2 months, corresponds to 1/3 to 1/6 of the full rotation period. Since there are two magnetic extrema every rotation period, the inclination of the rotation axis would have to be smaller than $\sim$10$^\circ$ to allow $\alpha$ to be $\leq 3.5^\circ$ over this whole rotation phase, with both $\beta$ (the angle between the rotation and the magnetic axis) and $\phi$ (the rotational phase) fine-tuned to minimise $\alpha$. It seems very improbable to have all these parameters conspire to hide a magnetic signature.

\section{Conclusions}
Although the presence of a magnetic field is a possible explanation for the asymmetry of the splitting of the triplet around 5.26 c\,d$^{-1}$, and the UV spectra show variability similar to what is observed in known magnetic stars, no magnetic field has been detected. $\nu$ Eri may still harbour a weak magnetic field, but it is highly unlikely that the observed pulsation mode splitting is the result of a 5--10 kG magnetic field.

In the absence of a magnetic field, we can conclude that the observed UV variability is due to the strong pulsations in this star, which is supported by the short timescale of the variability. However, the asymmetry of the splitting of the pulsation triplet around 5.26 c\,d$^{-1}$ remains unexplained. In view of the discovery of two more triplets with different splittings and asymmetries, more sophisticated modeling of this star and all three triplets is required before further conclusions can be drawn on the relation between the stellar rotation and the splitting, and asymmetry, of the triplets.

{\acknowledgements EV and HFH would like to thank W. Dziembowski for inspiring discussions on $\nu$ Eri during the Mmabatho meeting on magnetic fields in South Africa, November 2002 when this project was initiated. Most of the TBL observations were taken in service mode. Without this efficient observing mode this project could not have been done. In particular we acknowledge M. Auri\`ere and F. Paletou for observing. We are also indebted to the capable TBL staff for assisting with the observations, G. Handler for providing radial velocity information, and M. Smith, the referee, for his constructive comments. This
research has been partly based on INES data from the IUE satellite and made use of the Simbad and ADS 
databases maintained at CDS, Strasbourg, France.}

\bibliographystyle{aa}
\bibliography{../../references}

\end{document}